\documentclass[aps, prb, twocolumn, amsmath, amssymb, showpacs, superscriptaddress]{revtex4-1}

\usepackage[utf8]{inputenc}
\usepackage{graphicx}
\usepackage{units}
\usepackage{color}
\usepackage[pdftex, colorlinks=true, linkcolor=myblue, citecolor=myblue,
urlcolor=myblue]{hyperref}
\usepackage{times}

\definecolor{myblue}{rgb}{0,0,1}

\begin{document}

\title{Large current noise in nanoelectromechanical systems close to continuous
mechanical instabilities}

\author{Jochen Br\"uggemann}
\affiliation{Dahlem Center for Complex Quantum Systems \& Fachbereich Physik, 
Freie Universit\"at Berlin, 
Arnimallee 14, D-14195 Berlin, Germany}
\affiliation{I.\ Institut f\"ur Theoretische Physik, Universit\"at Hamburg,
Jungiusstra\ss e 9, D-20355 Hamburg, Germany}

\author{Guillaume Weick}
\email{Guillaume.Weick@ipcms.unistra.fr}
\affiliation{Institut de Physique et Chimie des Mat\'eriaux de Strasbourg,
Universit\'e de Strasbourg, CNRS UMR 7504, 23 rue du Loess, BP 43, F-67034
Strasbourg Cedex 2, France}

\author{Fabio Pistolesi}
\affiliation{Univ.\ Bordeaux, LOMA, UMR 5798, F-33400 Talence, France}
\affiliation{CNRS, LOMA, UMR 5798, F-33400 Talence, France}

\author{Felix von Oppen}
\affiliation{Dahlem Center for Complex Quantum Systems \& Fachbereich Physik, 
Freie Universit\"at Berlin, 
Arnimallee 14, D-14195 Berlin, Germany}


\begin{abstract}
We investigate the current noise of nanoelectromechanical systems close to
a continuous mechanical instability. In the vicinity of the latter, the
vibrational frequency of the nanomechanical system vanishes, rendering the system very 
sensitive to charge fluctuations  and, hence, resulting in very large
(super-Poissonian) current noise.
Specifically, we consider a suspended single-electron transistor close to the
Euler buckling instability. We show that such a system exhibits an exponential 
enhancement of the current
noise when approaching the Euler instability which we explain in terms of
telegraph noise. 
\end{abstract}

\pacs{73.63.-b, 85.85.+j, 63.22.Gh}

\maketitle

\section{Introduction}
\begin{figure}[tb]
\includegraphics[width=\columnwidth]{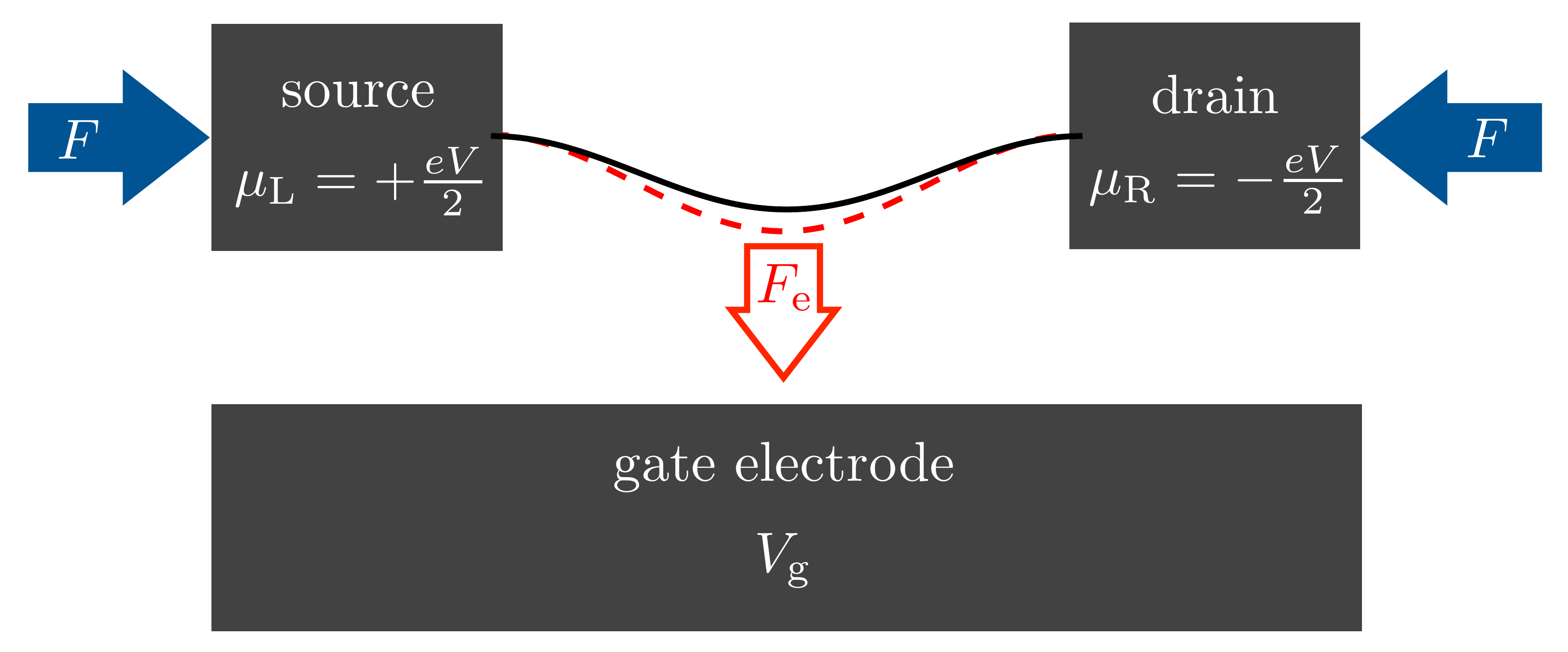}
\caption{\label{fig:setup}%
(Color online) Sketch of a suspended doubly clamped nanobeam 
forming a quantum dot (solid black line) connected 
via tunnel barriers to source and drain electrodes held at chemical 
potentials $\mu_\mathrm{L}$ and $\mu_\mathrm{R}$ by the bias voltage 
$V$. 
The lateral force $F$ compresses the nanobeam
and induces the buckling instability.
The beam is capacitively coupled to a metallic electrode 
kept at a gate voltage $V_\mathrm{g}$. This induces a stochastic force $F_\mathrm{e}$ that attracts the beam
towards the gate electrode whenever the quantum dot is charged (dashed red
line), inducing fluctuations of the nanobeam's deflection and,
in turn, fluctuations of the current through the device.}
\end{figure}

Nanoelectromechanical systems (NEMS) in which a nanomechanical resonator is
coupled to electronic degrees of freedom \cite{craig00_Science,
ekinc05_RSI, poot11_preprint} through, e.g., a single-electron transistor
(SET), show spectacular effects stemming from the coupling of the mechanical
part of the device to the electronic charge. These effects
arise due to the reduced size of the nanoresonator, so that
the backaction of the mechanical degree of freedom on the SET can have
significant consequences for the transport properties.
A prominent example
is the low-bias current blockade that occurs in the Coulomb blockade regime when the nanoresonator is
capacitively coupled to the SET. \cite{pisto07_PRB, pisto08_PRB} The presence of
an extra electron with charge $-e<0$ on the
central island forming a quantum dot on the suspended vibrating structure 
induces an electrostatic force $F_\mathrm{e}$ on the
resonator, shifting the equilibrium position of the latter by an amount
$F_\mathrm{e}/k$, with $k$ the spring constant of the oscillator (see Fig.~\ref{fig:setup}). This induces a
shift of the gate voltage $V_\mathrm{g}\sim F_\mathrm{e}^2/ek$ seen by the SET,
and, hence, a blockade of the current through the device for bias voltages
$V\lesssim F_\mathrm{e}^2/ek$.
This phenomenon is the classical counterpart of
the Franck-Condon blockade in molecular devices \cite{koch05_PRL, koch06_PRB}
that has recently been observed in carbon nanotube-based resonators for
high-energy longitudinal stretching modes.
\cite{letur09_NaturePhysics}
For classical nanoresonators, the current blockade has, to the best of our
knowledge, never been observed experimentally due to the relatively weak
electromechanical coupling $F_\mathrm{e}$ to the low-energy bending modes of the suspended
structure, although a precursor of this effect
has been lately reported in the literature. \cite{steel09_Science,
lassa09_Science}

It has been recently shown \cite{weick11_PRB, weick11b_PRB} how one can enhance 
the classical current blockade by orders of magnitude by exploiting the well-known 
Euler buckling instability. \cite{landau} Indeed, the spring constant
$k$ (or equivalently, the vibrational frequency of the
fundamental bending mode $\omega$) tends to zero when one brings
the nanoresonator to the buckling instability with the help of a lateral 
compression force $F$ (see Fig.\ \ref{fig:setup}).
Thus, the energy scale $F_\mathrm{e}^2/k$ at which the
current blockade occurs dramatically increases, rendering this phenomenon
potentially observable in future experiments. 

It is the purpose of the present paper to investigate the current noise in the
vicinity of a mechanical instability, using the Euler buckling instability as a
paradigmatic model.
We find that the current noise
(which contains valuable information about the dynamics of the nanomechanical
system
\cite{armou04_PRB, blant04_PRL, usman07_PRB,
pisto04_PRB, novot04_PRL, donar05_NJP, flind05_EPL})
is strongly enhanced in the vicinity of the Euler instability. 
The underlying source of noise arises from the stochastic nature of the charge
transfer processes. These are producing a current-induced stochastic force
acting on the mechanical degrees of freedom,
consisting of current-induced (conservative and nonconservative) averages as
well as a fluctuating force.
\cite{pisto07_PRB, pisto08_PRB, weick11_PRB, weick11b_PRB, blant04_PRL, mozyr06_PRB, 
husse10_PRB, nocer11_PRB, bode11_PRL, bode11_preprint, weick10_PRB}
Hence, the deflection of the nanotube exhibits a Langevin
dynamics which, due to the backaction of the nanoresonator on the SET, 
produces large super-Poissonian current noise. \cite{armou04_PRB, usman07_PRB,
pisto08_PRB, blant04_PRL, novot04_PRL, donar05_NJP, flind05_EPL} This effect is particularly strong
close to the Euler buckling
instability, where the nanoresonator becomes extremely soft ($k\to0$). 

Our theoretical approach employs
a nonequilibrium Born-Oppenheimer approximation, \cite{pisto07_PRB, pisto08_PRB,
blant04_PRL, mozyr06_PRB, husse10_PRB, nocer11_PRB, bode11_PRL, bode11_preprint}
which exploits the separation of timescales between fast electronic 
and slow mechanical degrees of freedom. 
This leads to an effective stochastic description of the nanoresonator 
in terms of a Langevin equation.
This approach becomes
asymptotically exact in the vicinity of the mechanical instability where
$\omega\to0$.
\cite{weick10_PRB, weick11_PRB, weick11b_PRB} 

The paper is organized as follows: Our model 
and the effective Langevin
description of the nanoresonator deflection is presented in
Sec.~\ref{sec:model}. In Sec.~\ref{sec:current}, we briefly recall the main
results of Ref.~\onlinecite{weick11_PRB} for the
current-voltage characteristics of the system 
that are essential for the understanding
of our numerical investigation of the current noise presented in Sec.~\ref{sec:noise}. 
In Sec.\ \ref{sec:thermal}, we detail the role played by thermal fluctuations on
the current noise and propose an analytical model based on telegraph noise that
reproduces most of our numerical findings. 
We present in Sec.\ \ref{sec:nonequilibrium} the role played by the full 
nonequilibrium fluctuations on the noise before we 
conclude in Sec.~\ref{sec:ccl}.

\section{Model}
\label{sec:model}
The model we adopt is the same as in Ref.\
\onlinecite{weick11_PRB} and 
we briefly recall it here for the convenience of the reader. 
The setup (see Fig.~\ref{fig:setup}) consists of a quantum dot embedded 
in a suspended nanobeam connected via tunnel barriers to source and drain
leads. A lateral compression force $F$ exerted on the nanobeam
brings it to the Euler buckling instability when $F$ approaches 
the critical force $F_\mathrm{c}$.
The gate electrode induces an electromechanical coupling between the
bending modes of the tube and the charge state of the dot proportional to the
electrostatic force $F_\mathrm{e}$. 
The Hamiltonian of
the system reads
\begin{equation}
\label{eq:H}
H=H_\mathrm{vib}+H_\mathrm{SET}+H_\mathrm{c}, 
\end{equation}
where $H_\mathrm{vib}$ describes the oscillating modes of the nanobeam,
$H_\mathrm{SET}$ the single-electron transistor, and $H_\mathrm{c}$ the coupling
between vibrational modes and electronic degrees of freedom.

At the Euler buckling instability ($F=F_\mathrm{c}$), the frequency of the fundamental
bending mode 
\begin{equation}
\label{eq:omega}
\omega=\omega_0\sqrt{-\delta}, \qquad \delta=\frac{F}{F_\mathrm{c}}-1
\end{equation}
vanishes, \cite{landau} $\omega_0$ being
the frequency of that mode for $F=0$, while all higher energy modes have a
finite frequency. At sufficiently low temperatures, we thus only consider the fundamental
bending mode and write \cite{weick10_PRB, weick11b_PRB, weick11_PRB, carr01_PRB,
werne04_EPL, peano06_NJP, savel06_NJP}
\begin{equation}
\label{eq:H_vib}
H_\mathrm{vib}=\frac{P^2}{2m}+\frac{m\omega^2}{2}X^2+\frac{\alpha}{4}X^4
\end{equation}
for the vibrational part of the Hamiltonian \eqref{eq:H}.
Here, $X$ is the deflection of the tube and $P$ its associated canonical
momentum with effective mass $m$. In Eq.\ \eqref{eq:H_vib}, the quartic term proportional to $\alpha>0$
ensures the stability of the system for $F>F_\mathrm{c}$ ($\omega^2<0$), where
the beam buckles into one of the two metastable positions at $X=\pm\sqrt{-m\omega^2/\alpha}$.
For $F<F_\mathrm{c}$ ($\omega^2>0$), the beam remains flat. 

We model the SET by a single-level quantum dot with orbital energy
$\epsilon_\mathrm{d}$ connected via tunnel barriers to left (L) and right (R) leads
held at chemical potential $\mu_\mathrm{L}$ and $\mu_\mathrm{R}$, respectively.
The SET Hamiltonian reads 
$H_\mathrm{SET}=H_\mathrm{dot}+H_\mathrm{leads}+H_\mathrm{tun}$,
where the dot Hamiltonian
$H_\mathrm{dot}=(\epsilon_\mathrm{d}-e\bar V_\mathrm{g})n_\mathrm{d}+U
n_\mathrm{d} (n_\mathrm{d}-1)/2$
is expressed in terms of $n_\mathrm{d}=d^\dagger d$, 
$d$ annihilating an electron on the dot. Here, $\bar
V_\mathrm{g}$ is the (effective) applied
gate voltage.
The intradot Coulomb repulsion is denoted by $U$
and is assumed to be the largest energy scale in the problem, thus preventing
double occupancy of the dot. 
The Hamiltonian for the (spinless) electrons with
energy $\epsilon_k$ and momentum $k$ in
the two leads (annihilated by the operator $c_{ka}$, $a=\mathrm{L,R}$) is written as 
$H_\mathrm{leads}=\sum_{ka}(\epsilon_k-\mu_a)c_{ka}^\dagger
c_{ka}^{\phantom{\dagger}}$.
Finally,
tunneling between dot and leads is accounted for by the Hamiltonian 
$H_\mathrm{tun}=\sum_{ka}(t_ac_{ka}^\dagger d+\mathrm{h.c.})$,
with $t_a$ the
tunneling amplitude between the dot and lead $a$. In the remainder of this
paper, we assume the temperature $T$ to be much larger than the
tunneling-induced width $\Gamma=\sum_a\Gamma_a$ of the dot orbital
(sequential tunneling regime). This weak-coupling assumption should not
qualitatively change our results for the low-frequency current noise, as it is
the case for the $I$-$V$ characteristics which is qualitatively the same in the
sequential \cite{weick11_PRB} and cotunneling \cite{weick11b_PRB} regimes. 
Moreover, we consider for simplicity symmetric voltage
drops ($\mu_\mathrm{L}=-\mu_\mathrm{R}=eV/2$) and symmetric coupling to the
leads ($\Gamma_\mathrm{L}=\Gamma_\mathrm{R}=\Gamma/2$).

The coupling Hamiltonian between vibrational and electronic degrees of freedom 
entering Eq.~\eqref{eq:H}, 
\begin{equation}
\label{eq:H_c}
H_\mathrm{c}=F_\mathrm{e}Xn_\mathrm{d},
\end{equation}
arises due to the capacitive coupling between the gate electrode and the
nanobeam when the latter is charged with one extra electron. \cite{sapma03_PRB,
steel09_Science, lassa09_Science} Since the dot occupation $n_\mathrm{d}$ is
a stochastic variable (taking values 0 and 1 in our model), the coupling
\eqref{eq:H_c} produces a random electrostatic force with magnitude 
$F_\mathrm{e}$ on the nanobeam (see Fig.~\ref{fig:setup}). 

As $\omega\to0$ close to the buckling instability, 
the oscillator becomes classical ($\hbar\omega\ll
k_\mathrm{B}T$) and slow compared to the electronic degrees of freedom
($\omega\ll\Gamma$). This justifies a nonequilibrium Born-Oppenheimer
approximation,
\cite{blant04_PRL, weick10_PRB, weick11_PRB, weick11b_PRB, mozyr06_PRB, 
husse10_PRB, nocer11_PRB, bode11_PRL, bode11_preprint} in which
the vibrational dynamics is characterized
by a Langevin process with white noise, 
\begin{equation}
\label{eq:Langevin_scaled}
\frac{\mathrm{d}^2x}{\mathrm{d}\tau^2}
+\left[\gamma(x)+\gamma_\mathrm{e}\right]\frac{\mathrm{d}x}{\mathrm{d}\tau}
=f_\mathrm{eff}(x)+\xi(x, \tau).
\end{equation}
Here and in what follows, we use reduced units 
$x=X/\ell$, $p=P/m\omega_0\ell$ and $\tau=\omega_0 t$ in terms of the polaron
shift $\ell=F_\mathrm{e}/m\omega_0^2$ and the relevant energy scale of the
problem
$E_\mathrm{E}^0=F_\mathrm{e}\ell$ (for more details, see Ref.\
\onlinecite{weick11_PRB}).
In Eq.~\eqref{eq:Langevin_scaled}, the effective force acting on the nanobeam,
\begin{equation}
\label{eq:f_eff}
f_\mathrm{eff}(x)=\delta x-\tilde\alpha x^3-n_0(x), 
\end{equation}
with $\tilde\alpha=\alpha \ell^4/E_\mathrm{E}^0$,
arises (i) from the bare vibrational Hamiltonian \eqref{eq:H_vib} and (ii) from
the coupling between vibrational and electronic degrees of freedom
\eqref{eq:H_c}. This current-induced force is proportional to the
occupation of the dot for fixed $x$, which, in the sequential tunneling
regime ($\hbar\Gamma\ll k_\mathrm{B}T$), is given by
\begin{equation}
\label{eq:n0_reduced}
n_0(x)=\frac 12 \left[f_\mathrm{L}(x)
+f_\mathrm{R}(x)\right],
\end{equation}
with
\begin{equation}
\label{eq:Fermi}
f_\mathrm{L/R}(x)=\left[\exp{\left(\frac{x-v_\mathrm{g}\mp v/2}{\tilde
T}\right)}+1\right]^{-1}
\end{equation}
the
Fermi function in the left and right leads, respectively.
Here we introduced a reduced bias voltage 
$v=eV/E_\mathrm{E}^0$, gate voltage $v_\mathrm{g}=(e\bar
V_\mathrm{g}-\epsilon_\mathrm{d})/E_\mathrm{E}^0$,
and temperature $\tilde T=k_\mathrm{B}T/E_\mathrm{E}^0$.
The charge fluctuations on the quantum dot
induce a fluctuating force $\xi(x, \tau)$ in the Langevin equation
\eqref{eq:Langevin_scaled}, with average $\langle \xi(x, \tau)\rangle=0$ and white-noise correlator 
$\langle\xi(x, \tau)\xi(x, \tau')\rangle=[d(x)+2\gamma_\mathrm{e}
\tilde T]\delta(\tau-\tau')$. 
Here, 
the current-induced fluctuation is given by \cite{pisto07_PRB, weick11_PRB}
\begin{equation}
\label{eq:d}
d(x)=\frac{2\omega_0}{\Gamma}n_0(x)\left[1-n_0(x)\right], 
\end{equation}
and the extrinsic damping constant $\gamma_\mathrm{e}$ accounts for the finite
quality factor $Q=1/\gamma_\mathrm{e}$ of the nanoresonator.
Finally, retardation effects in the response of the resonator to the current
flow lead to a current-induced dissipative force
with friction coefficient \cite{pisto07_PRB, weick11_PRB}
\begin{equation}
\label{eq:gamma}
\gamma(x)=-\frac{\omega_0}{\Gamma}\frac{\partial}{\partial x}n_0(x).
\end{equation}

The Langevin
equation \eqref{eq:Langevin_scaled} is equivalent to the Fokker-Planck equation
\cite{zwanzig}
\begin{equation}
\label{eq:FP_scaled}
\frac{\partial}{\partial \tau}\mathcal{P}(x, p, \tau)=
{\mathcal{L}}\mathcal{P}(x, p, \tau)
\end{equation}
for the probability distribution $\mathcal{P}(x, p, \tau)$ that the nanobeam is at phase-space
point $(x,p)$ at time $\tau$. 
In Eq.\ \eqref{eq:FP_scaled}, the Fokker-Planck operator is defined as 
\begin{align}
\label{eq:L}
{\mathcal{L}}=&
-p\frac{\partial}{\partial x}-f_\mathrm{eff}(x)\frac{\partial}{\partial p} 
+\left[\gamma(x)+\gamma_\mathrm{e}\right]\frac{\partial}{\partial p}p
\nonumber\\
&+\left[\frac{d(x)}{2}+\gamma_\mathrm{e}\tilde
T\right]\frac{\partial^2}{\partial p^2}.
\end{align}
Solving the Fokker-Planck equation \eqref{eq:FP_scaled} [or equivalently the Langevin
equation \eqref{eq:Langevin_scaled}] gives access to both the dynamics of the
vibrational mode of the nanoresonator and the resulting transport properties of
the device, such as its $I$-$V$ characteristics (see Sec.~\ref{sec:current}) 
and its noise power spectrum (see Secs.~\ref{sec:noise}, \ref{sec:thermal} and
\ref{sec:nonequilibrium}).

\section{Current blockade}
\label{sec:current}
Due to the separation of timescales between slow vibrational motion and fast
electronic dynamics, the average current 
\begin{equation}
\label{eq:I}
I=\int\mathrm{d}x\mathrm{d}p\;\mathcal{P}_\mathrm{st}(x, p)\mathcal{I}(x)
\end{equation}
is obtained by averaging the quasistationary current 
\begin{equation}
\label{eq:I(x)}
\mathcal{I}(x)=\frac{e\Gamma}{4}
\left[
f_\mathrm{L}(x)-f_\mathrm{R}(x)
\right]
\end{equation}
for fixed deflection $x$ over the 
stationary solution $\mathcal{P}_\mathrm{st}$ of the Fokker-Planck equation
\eqref{eq:FP_scaled}.

\begin{figure*}[tbh]
\includegraphics[width=\linewidth]{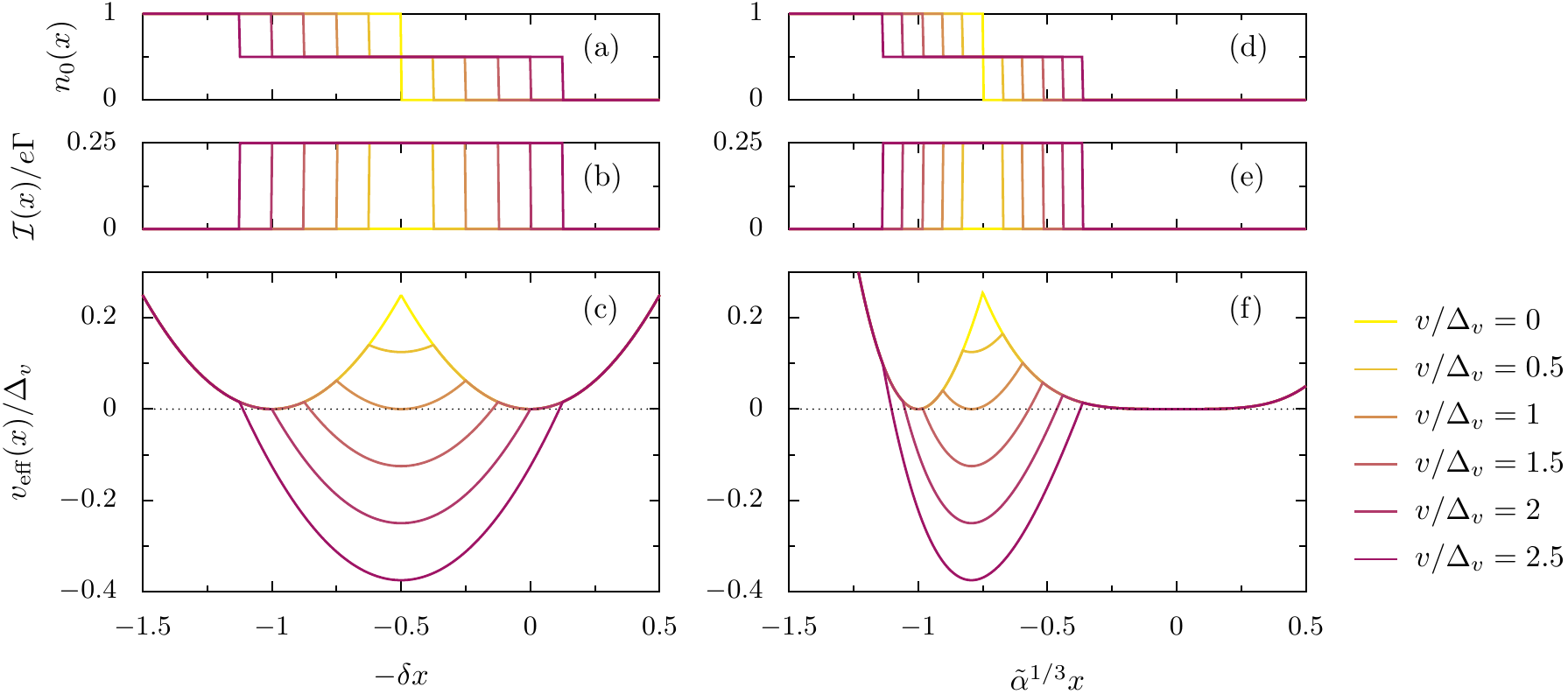}
\caption{\label{fig:v_eff}%
(Color online)
Zero temperature
(a,d) average occupation of the dot for fixed $x$ [Eq.\ \eqref{eq:n0_reduced}], 
(b,e) quasistationary current [Eq.\ \eqref{eq:I(x)}] and
(c,f) effective potential [Eq.\ \eqref{eq:v_eff}], (a,b,c) far below the Euler instability
($-\delta\gg\tilde\alpha^{1/3}$) and (d,e,f) at the instability ($\delta=0$) for
increasing bias voltages and for a gate voltage
$v_\mathrm{g}=v_\mathrm{g}^\mathrm{min}$ [cf.\ Eq.\ \eqref{eq:vg}].}
\end{figure*}

The classical current blockade phenomenon \cite{weick11_PRB, pisto07_PRB} can be understood in
terms of the effective potential
\begin{equation}
\label{eq:v_eff}
v_\mathrm{eff}(x)
=-\frac{\delta x^2}{2}+\frac{\tilde\alpha x^4}{4}+x
+\frac{\tilde T}{2}
\ln{\big(
f_\mathrm{L}(x)f_\mathrm{R}(x)\big)}
\end{equation}
associated to the effective force \eqref{eq:f_eff}.
The effective potential is shown in Fig.\ \ref{fig:v_eff} for $\tilde T=0$
together with
the average occupation of the dot \eqref{eq:n0_reduced} and the quasistationary current
\eqref{eq:I(x)} for compression forces
corresponding to the beam far below
[Figs.\ \ref{fig:v_eff}(a)--\ref{fig:v_eff}(c)] and at the Euler 
instability [Figs.\ \ref{fig:v_eff}(d)--\ref{fig:v_eff}(f)]. 
\cite{footnote:v_eff} In both cases, the most stable minima of
$v_\mathrm{eff}(x)$ correspond, for bias voltages $v$ smaller than the gap
\cite{weick11_PRB}
\begin{equation}
\label{eq:gap}
\Delta_v=
\begin{cases}
-1/2\delta, & -\delta\gg{\tilde\alpha}^{1/3},\\
1/4\delta, & \delta\gg{\tilde\alpha}^{1/3},\\
\frac{2^{1/3}-1}{{\tilde\alpha}^{1/3}}
\left(
\frac{3}{2^{4/3}}
-\frac{\delta}{\tilde\alpha^{1/3}}
\right),
& |\delta|\ll{\tilde\alpha}^{1/3},
\end{cases}
\end{equation}
to an average
occupation $n_0(x)=0$ or $1$, i.e., the system is not conducting [``blocked"
minima for which $\mathcal{I}(x)=0$, cf.\ Figs.\ \ref{fig:v_eff}(b) and \ref{fig:v_eff}(e)]. 
For $v>\Delta_v$, the most stable minimum corresponds to $n_0(x)=1/2$
and the current can flow [``conducting" minimum corresponding to
$\mathcal{I}(x)\neq0$]. At the threshold $v=\Delta_v$, the
three minima are metastable, leading to a current-induced instability of the system.
Since for relevant experimental parameters, $\tilde\alpha\ll1$,
\cite{steel09_Science, lassa09_Science, weick11_PRB} 
the gap \eqref{eq:gap} is maximal at the instability where $\delta=0$
[$F=F_\mathrm{c}$, cf.\ Eq.\ \eqref{eq:omega}] and is orders of
magnitude larger than for $\delta=-1$ ($F=0$).
The gaps of Eq.~\eqref{eq:gap} are obtained for gate
voltages $v_\mathrm{g}=v_\mathrm{g}^\mathrm{min}$, with
\cite{weick11_PRB}
\begin{equation}
\label{eq:vg}
v_\mathrm{g}^\mathrm{min}=
\begin{cases}
1/2\delta, & -\delta\gg{\tilde\alpha}^{1/3},\\
-1/4\delta-\sqrt{\delta/\tilde\alpha}, &
\delta\gg{\tilde\alpha}^{1/3},\\
-\frac{1}{4{\tilde\alpha}^{1/3}}
\left(
3 
+\frac{2\delta}{\tilde\alpha^{1/3}}
\right),
& |\delta|\ll{\tilde\alpha}^{1/3}.
\end{cases}
\end{equation}

\begin{figure*}[tbh]
\includegraphics[width=\linewidth]{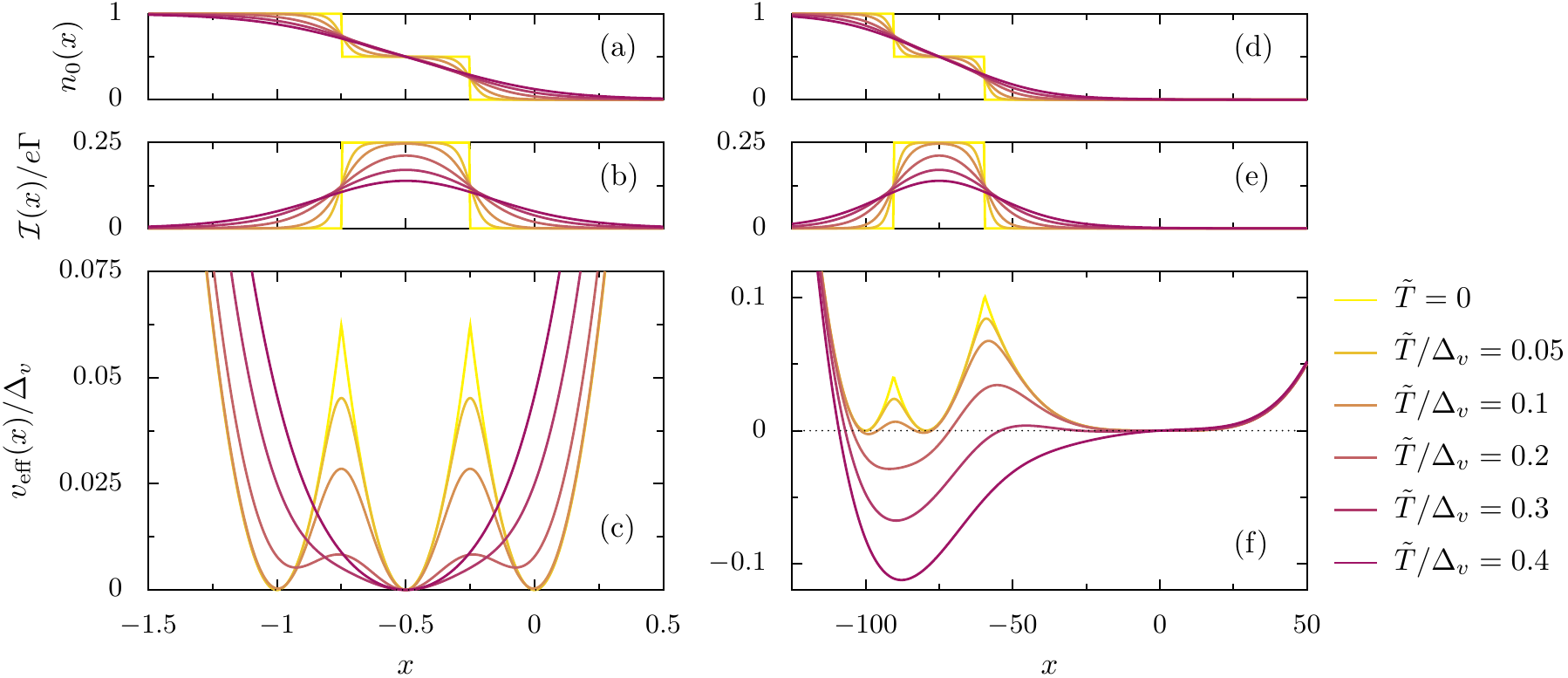}
\caption{\label{fig:v_eff_finite_T}%
(Color online)
For increasing temperature, 
(a,d) average occupation of the dot for fixed $x$, 
(b,e) quasistationary current 
and (c,f)
effective potential, (a,b,c) far below the Euler instability
($\delta=-1$) and (d,e,f) at the instability ($\delta=0$).
In the figure, $v=\Delta_v$, 
$v_\mathrm{g}=v_\mathrm{g}^\mathrm{min}$
and $\tilde\alpha=10^{-6}$.}
\end{figure*}

At finite temperatures, the effective potential \eqref{eq:v_eff} qualitatively
changes (see Fig.\ \ref{fig:v_eff_finite_T}): the barriers separating the minima 
are lowered as one increases the temperature, 
leaving the effective potential with a single
minimum for large enough temperatures.
As detailed in Ref.\ \onlinecite{weick11_PRB}, the current blockade becomes less
pronounced as the temperature increases [see Fig.\ \ref{fig:temperature}(a)
below].
Moreover, taking into account the current-induced fluctuations \eqref{eq:d} and
dissipation \eqref{eq:gamma} in the Langevin dynamics \eqref{eq:Langevin_scaled}
further reduces the current blockade [see Fig.\ \ref{fig:nonequilibrium}(a)
below].

\section{Current noise at the Euler buckling instability}
\label{sec:noise}
We start by discussing the two contributions to the current noise, i.e., the usual shot
noise and
the mechanically-induced noise. 
It is the latter contribution to the noise which we find to be 
dramatically enhanced close to the mechanical instability.

The noise power spectrum is defined as \cite{blant00_PhysRep}
\begin{equation}
\label{eq:S}
S(\Omega)=2\int_{-\infty}^{+\infty}\mathrm{d}t\;\mathrm{e}^{\mathrm{i}\Omega t}
\langle \Delta \hat I(t_0+t)\Delta \hat I(t_0)\rangle, 
\end{equation}
where $\Delta \hat I(t)=\hat I(t)-\langle\hat I\rangle$ denotes the
time-dependent fluctuations of the current operator.
In Eq.\ \eqref{eq:S}, the brackets $\langle\ldots\rangle$ indicate an ensemble average
or, equivalently, an average over the initial time $t_0$.
Due to the separation of timescales between fast electronic dynamics and slow
vibrational motion ($\omega\ll\Gamma$), one can identify two additive
contributions to the noise power spectrum \eqref{eq:S},
$S=S_\mathrm{sh}+S_\mathrm{m}$. \cite{pisto08_PRB,
blant04_PRL, mozyr06_PRB, usman07_PRB} The first one, $S_\mathrm{sh}$, corresponds to the
(thermal) Nyquist-Johnson and shot noise and is discussed in Appendix
\ref{sec:shot}.
The second contribution to Eq.\ \eqref{eq:S}, the mechanically-induced
noise 
$S_\mathrm{m}$ (referred to as ``mechanical noise" in the sequel), 
is induced by the fluctuations of the nanobeam deflection $x(t)$.
These occur on a much longer timescale (of the order of $1/\omega$) than the 
shot noise (the corresponding current-current correlator decaying in that case
on the short timescale $1/\Gamma$). The mechanically-induced noise therefore
dominates the noise power spectrum at low frequencies, and can exceed the shot
noise by orders of magnitude. \cite{pisto08_PRB, blant04_PRL, usman07_PRB,
mozyr06_PRB} 

The mechanical noise reads
\begin{align}
\label{eq:S_m}
S_\mathrm{m}(\Omega)=&\;2\int
\frac{\mathrm{d}\tau}{\omega_0}\;\mathrm{e}^{\mathrm{i}\Omega \tau/\omega_0}
\int
\mathrm{d}x
\mathrm{d}p
\mathrm{d}x_0
\mathrm{d}p_0
\;
\Delta \mathcal{I}(x)
\nonumber\\
&\times
\mathcal{P}(x, p, \tau| x_0, p_0, \tau_0)
\Delta \mathcal{I}(x_0)
\mathcal{P}_\mathrm{st}(x_0, p_0)
\end{align}
where $\mathcal{P}(x, p, \tau| x_0, p_0, \tau_0)$ is the conditional probability that the nanobeam
is at phase-space point $(x, p)$ at time $\tau$, provided it was at 
$(x_0, p_0)$ at time $\tau_0\equiv0$. 
In Eq.~\eqref{eq:S_m},  $\Delta \mathcal{I}(x)=\mathcal{I}(x)-I$ is the quasistationary
current fluctuation, with $\mathcal{I}(x)$ and $I$ given by Eqs.~\eqref{eq:I(x)}
and \eqref{eq:I}, respectively. The conditional probability $\mathcal{P}(x, p,
\tau| x_0, p_0, \tau_0)$ can be obtained from the
time-dependent solution of the Fokker-Planck equation \eqref{eq:FP_scaled} with
the initial condition $\mathcal{P}(x, p, \tau_0|x_0, p_0,
\tau_0)=\delta(x-x_0)\delta(p-p_0)$.
Equation \eqref{eq:S_m} can then be re-expressed by exploiting the
above initial condition and performing the Laplace transform of the
Fokker-Planck equation \eqref{eq:FP_scaled}. This procedure yields \cite{pisto08_PRB}
\begin{align}
\label{eq:S_transformed}
S_\mathrm{m}(\Omega)=&-\frac{4}{\omega_0}\int \mathrm{d}x\mathrm{d}p\;
\Delta\mathcal{I}(x)
\nonumber\\
&\times\left[{\mathcal{L}}^2+\left(\frac{\Omega}{\omega_0}\right)^2\right]^{-1}
{\mathcal{L}}\left[\Delta \mathcal{I}(x)\mathcal{P}_\mathrm{st}(x,
p)\right], 
\end{align}
with ${\mathcal{L}}$ the Fokker-Planck operator defined in Eq.\
\eqref{eq:L}. \cite{flind05_PhysicaE}

In the form of Eq.\ \eqref{eq:S_transformed}, the mechanical noise can be 
straightforwardly computed numerically, since it only requires the knowledge of
the stationary probability distribution $\mathcal{P}_\mathrm{st}$ corresponding
to the Fokker-Planck equation \eqref{eq:FP_scaled}, as it is the case for the average
current \eqref{eq:I} (for details, see Ref.\
\onlinecite{pisto08_PRB}). \cite{footnote:numerics}
Alternatively, one can solve for the time-dependent solution $x(\tau)$ of the Langevin equation \eqref{eq:Langevin_scaled}
using standard techniques for stochastic differential equations. \cite{kloeden} The average current and noise are 
then computed by performing the time averages of the quasistationary current
\eqref{eq:I(x)} and the current-current correlator entering Eq.\ \eqref{eq:S},
respectively. We have checked numerically that this approach
yields the same results as the
ones based on the stationary solution of the Fokker-Planck equation
\eqref{eq:FP_scaled} [see Eqs. \eqref{eq:I} and \eqref{eq:S_transformed}]. 
However, although more physically transparent, the method based on Eq.\
\eqref{eq:Langevin_scaled} requires long simulation 
times as well as sampling. We thus 
use the other method for all the numerical results 
presented in the sequel of the paper [except in Fig.\ \ref{fig:Langevin},
where we explicitly simulate the time-dependent deflection of the nanobeam 
from the Langevin equation \eqref{eq:Langevin_scaled}]. 

\begin{figure}[tb]
\includegraphics[width=\columnwidth]{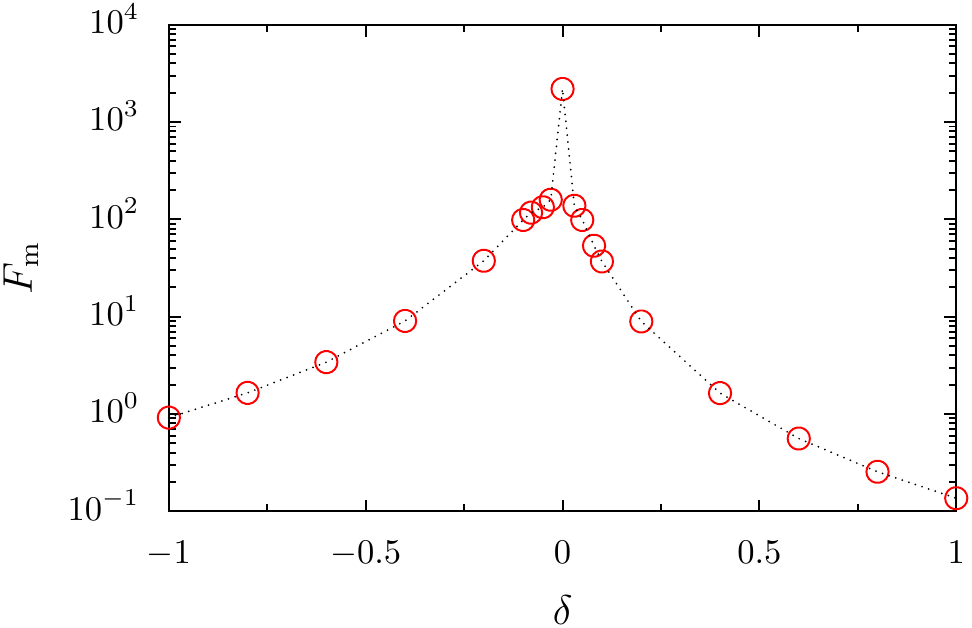}
\caption{\label{fig:main_result}%
(Color online) 
Fano factor as a function of the reduced
compression force $\delta$ [cf.\ Eq.\ \eqref{eq:omega}]. In the figure,
$\gamma_\mathrm{e}=\omega_0/\Gamma=10^{-2}$, $\tilde T=3$, $\tilde\alpha=10^{-6}$,
$v=\Delta_v$ and
$v_\mathrm{g}=v_\mathrm{g}^\mathrm{min}$.
The red circles correspond to the data points, while the dotted line serves as a guide to the eye.}
\end{figure}

In Fig.\ \ref{fig:main_result}, we present our numerical results for the Fano factor 
$F_\mathrm{m}=S_\mathrm{m}(0)/2e|I|$, 
where the zero-frequency noise $S_\mathrm{m}(0)$ and the average current $I$,
Eqs.\ \eqref{eq:S_transformed} and \eqref{eq:I}, respectively, are computed
for typical parameters as a function of the reduced force
$\delta$. In Fig.\ \ref{fig:main_result}, the bias and gate
voltages correspond to the apex of the Coulomb diamond [$v=\Delta_v$ and
$v_\mathrm{g}=v_\mathrm{g}^\mathrm{min}$, cf.\ Eqs.\ \eqref{eq:gap} and
\eqref{eq:vg}, respectively]. As envisioned above, there is a dramatic
increase of the mechanically-induced current noise in the vicinity of the Euler
buckling instability ($\delta\approx0$ in Fig.\ \ref{fig:main_result}) as
compared to the noise far away from the instability. Moreover, 
the Fano factor can take, depending on the compression force $\delta$, super-Poissonian values ($F_\mathrm{m}>1$) that are well above the shot
noise contribution, $F_\mathrm{sh}=1/2$ (see Appendix \ref{sec:shot}).

The results of Fig.\ \ref{fig:main_result} can essentially be understood in
terms of telegraph noise in the effective potential \eqref{eq:v_eff} (see also
Figs.\ \ref{fig:v_eff} and \ref{fig:v_eff_finite_T}). \cite{footnote:shuttle}
Indeed, unlike the energy gap \eqref{eq:gap} which
varies algebraically with the force $\delta$ as $\sim 1/|\delta|$, the numerical results of Fig.\
\ref{fig:main_result} indicate that the noise (or Fano factor) depends
\textit{exponentially} on $1/|\delta|$ (notice the logarithmic scale in Fig.\
\ref{fig:main_result}), suggesting telegraph noise. 
As the compression force increases towards the
instability at $\delta=0$, the height of the barriers separating the three
metastable minima (for $v=\Delta_v$) grows as the gap \eqref{eq:gap}, such
that the waiting time of the system in one of these minima increases exponentially. Thus,
the probability for the system to switch to another minimum is drastically reduced, 
subsequently increasing the telegraph noise. As the height of the potential barriers near
the Euler instability is very large, scaling as $1/\tilde\alpha^{1/3}$
with $\tilde\alpha\ll1$ [see Fig.\ \ref{fig:v_eff}(f)], and the energy gap
\eqref{eq:gap} is maximal at the instability, this leads to a current noise
which is also maximal at the Euler instability.

\begin{figure}[tb]
\includegraphics[width=1\columnwidth]{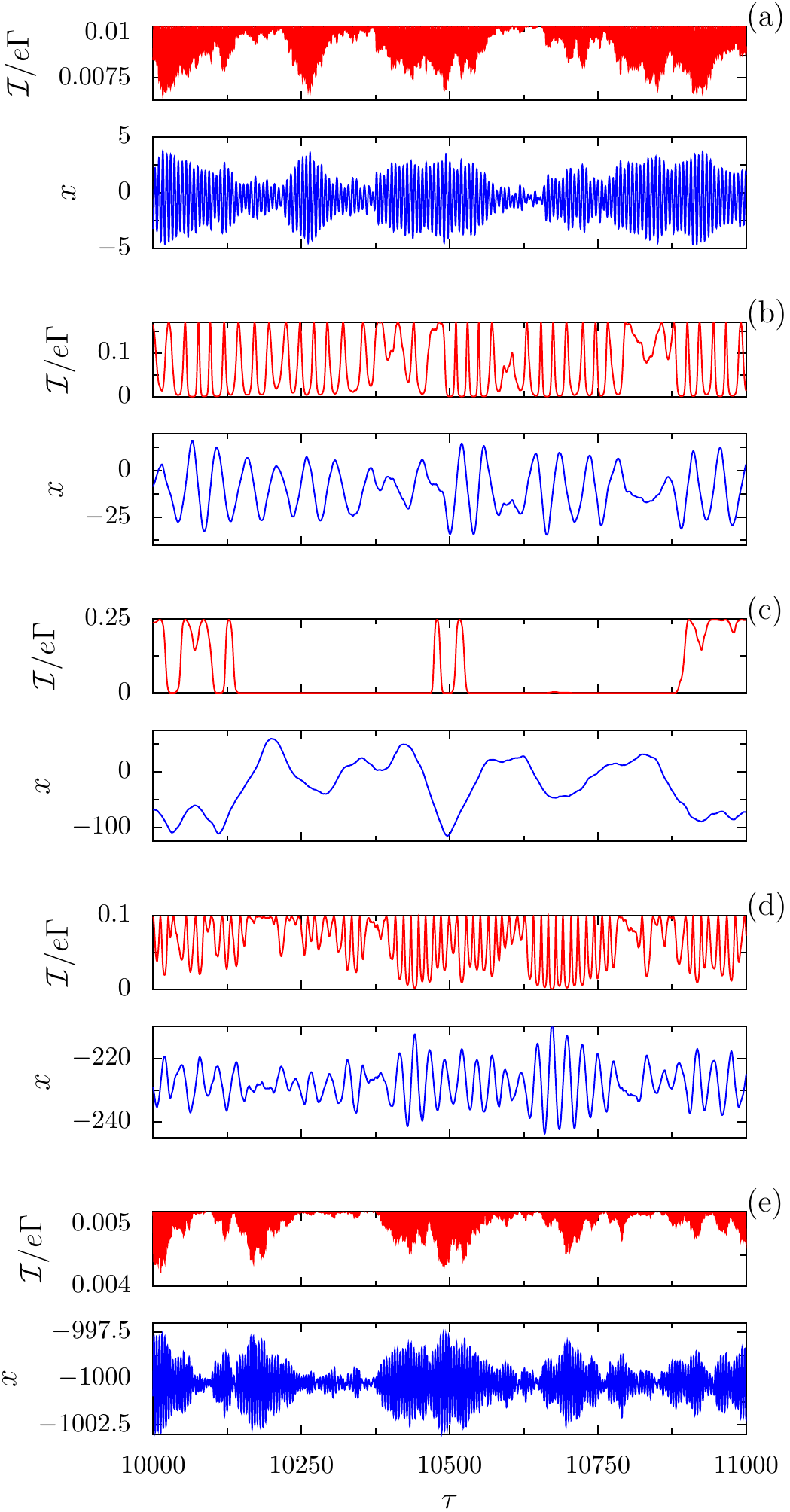}
\caption{\label{fig:Langevin}%
(Color online) 
Deflection $x$ (in blue) and quasistationary current $\mathcal I$ (in red) as
a function of time simulated by the Langevin equation
\eqref{eq:Langevin_scaled}.
In the figure,
the compression force increases from (a) to (e): 
(a) $\delta=-1$, 
(b) $\delta=-0.05$,
(c) $\delta=0$,
(d) $\delta=0.05$, and
(e) $\delta=1$.
The parameters
used in the figure 
are the same as in Fig.\ \ref{fig:main_result}.}
\end{figure}

This interpretation is confirmed by Fig.\ \ref{fig:Langevin}, which shows the result of a 
simulation \cite{kloeden} of the deflection of the nanobeam
$x$ as a function of time (see blue lines in the figure) as obtained from the Langevin equation \eqref{eq:Langevin_scaled} for
the same parameters as in Fig.\ \ref{fig:main_result}. We also show the
resulting quasistationary current \eqref{eq:I(x)} as a function of time by red
lines in Fig.\
\ref{fig:Langevin}. 
Far from the instability [Figs.\ \ref{fig:Langevin}(a) and
\ref{fig:Langevin}(e)], the dynamics of the nanobeam follows qualitatively the
behavior of a Brownian particle in a harmonic potential. Indeed, for the
temperature used in Figs.\ \ref{fig:main_result} and \ref{fig:Langevin}, the
effective potential \eqref{eq:v_eff} far from the instability has a single
minimum [see also Fig.\ \ref{fig:v_eff_finite_T}(c)]. For temperatures $\tilde T$
which are large compared to the gap \eqref{eq:gap}, the current shown
in Figs.\ \ref{fig:Langevin}(a) and \ref{fig:Langevin}(e) switches 
rapidly between values which are small as compared to the maximal current
$e\Gamma/4$ [cf.\ Fig.\ \ref{fig:v_eff_finite_T}(b)]. Hence, the resulting Fano factor is relatively small
(cf.\ Fig.\ \ref{fig:main_result}). As one approaches the
Euler instability from below [Fig.\ \ref{fig:Langevin}(b)] or above [Fig.\
\ref{fig:Langevin}(d)], the dynamics of the nanobeam becomes slower.
Then the
behavior of the current as a function of time starts to resemble telegraph
noise, as the effective potential starts developing metastable minima for this
value of the temperature as compared to the energy gap \eqref{eq:gap}.
At the instability [Fig.\ \ref{fig:Langevin}(c)], the dynamics of the nanobeam
becomes very slow, and
the behavior of the current as a function of time is completely stochastic and
uncorrelated, with long waiting times between vanishing
and maximal current. The corresponding Fano factor is thus extremely large and
super-Poissonian ($F_\mathrm{m}>10^3$ in Fig.\ \ref{fig:main_result}), and much
larger than far from the Euler instability.

In order to understand the features of our main numerical
results presented in Figs.\
\ref{fig:main_result} and \ref{fig:Langevin} in more detail, we will first consider the role played by
thermal fluctuations alone (Sec.\ \ref{sec:thermal}), while the full
nonequilibrium dynamics is investigated further in Sec.\
\ref{sec:nonequilibrium}.

\section{Thermal fluctuations}
\label{sec:thermal}
In this section, we consider the fully adiabatic limit
$\omega_0/\Gamma\rightarrow0$. We neglect the current-induced fluctuations
and dissipation in the Fokker-Planck equation \eqref{eq:FP_scaled} [cf.\
Eqs.~\eqref{eq:d} and \eqref{eq:gamma}] and consider the role played by thermal
fluctuations alone. We start in Sec.\ \ref{sec:telegraph} with a simplified
analytical model based on standard telegraph noise. In Sec.\ 
\ref{sec:numerics} we substantiate our
findings by evaluating Eq.\
\eqref{eq:S_transformed} numerically.

\subsection{Telegraph noise}
\label{sec:telegraph}
We present here an analytical estimate of the noise power spectrum.
Our simplified model relies on thermally-induced telegraph noise
\cite{machl54_JAP} and on an estimate of the
escape rates based on Kramers reaction rate theory. \cite{zwanzig,
krame40_Physica, hangg90_RMP}

In what follows, we work in the low-temperature regime $\tilde T\ll\Delta_v$,
with the gap $\Delta_v$ given in Eq.\ \eqref{eq:gap}. Moreover, we focus on the
case where the nanobeam is far below the Euler instability
($-\delta\gg\tilde\alpha^{1/3}$), as the results presented below should stay
qualitatively the same for larger compression forces. We thus approximate the effective potential
\eqref{eq:v_eff}
by its zero-temperature counterpart, which is
shown in Fig.\ \ref{fig:v_eff}(c) for a gate voltage
$v_\mathrm{g}=v_\mathrm{g}^\mathrm{min}$ [cf.\ Eq.\ \eqref{eq:vg}]. 
As one can see from Fig.\ \ref{fig:v_eff}(c), the effective
potential has three metastable minima
for $0<v<2\Delta_v$: two of them are equivalent (located symmetrically at
$x_1=1/\delta$ and $x_0=0$ about the line
$x=1/2\delta$) and correspond
to a state in which the current vanishes [see Fig.\ \ref{fig:v_eff}(b)], while
the one at $x_{1/2}=1/2\delta$ corresponds to a current-carrying state. 
This suggests to write a rate equation for the probabilities $P_\mathrm{c}$ and
$P_\mathrm{b}$ that the system is in a conducting or blocked state,
respectively. Denoting $\Gamma_\mathrm{in}=(\Gamma_{x_0\rightarrow
x_{1/2}}+\Gamma_{x_1\rightarrow x_{1/2}})/2$ and
$\Gamma_\mathrm{out}=\Gamma_{x_{1/2}\rightarrow x_1}+\Gamma_{x_{1/2}\rightarrow x_0}$ the
transition rates in and out of the conducting state ($\Gamma_{x_i\rightarrow
x_j}$ is here the transition rate from the minimum located at $x_i$ to the one
at $x_j$), we have 
$\dot P_\mathrm{c}=-\dot
P_\mathrm{b}=-\Gamma_\mathrm{out}P_\mathrm{c}+\Gamma_\mathrm{in}P_\mathrm{b}$.
Following Ref.\ \onlinecite{machl54_JAP}, the average current and the noise power
spectrum are readily obtained from the above rate equation. They read
\begin{equation}
\label{eq:I_approx}
I=\frac{e\Gamma}{4}\frac{\Gamma_\mathrm{in}}{\Gamma_\mathrm{in}+\Gamma_\mathrm{out}}
\end{equation}
and 
\begin{equation}
\label{eq:S_approx}
S_\mathrm{m}(\Omega)=\frac{e^2\Gamma^2}{4}
\frac{\Gamma_\mathrm{in}\Gamma_\mathrm{out}}
{\Gamma_\mathrm{in}+\Gamma_\mathrm{out}}
\frac{1}{\Omega^2+(\Gamma_\mathrm{in}+\Gamma_\mathrm{out})^2}, 
\end{equation}
respectively.
Notice that for bias voltages $v\geqslant2\Delta_v$, the effective potential
\eqref{eq:v_eff} has a single
minimum [see Fig.\ \ref{fig:v_eff}(c)], and the telegraph noise model presented above 
does not apply. Instead, the system's dynamics is characterized in that case by
standard Brownian noise.

As detailed in Appendix \ref{sec:rates}, the transition rates entering Eqs.\
\eqref{eq:I_approx} and \eqref{eq:S_approx}
can be easily calculated using
Kramers theory. \cite{krame40_Physica, hangg90_RMP, zwanzig}
Incorporating Eq.\ \eqref{eq:rates} in Eq.\ \eqref{eq:I_approx},
we find for the average current the approximate expression
\begin{equation}
\label{eq:current_telegraph}
I=\frac{e\Gamma}{4}\left\{1+2\exp{\left[\frac{\Delta_v}{4\tilde
T}\left(1-\frac{v}{\Delta_v}\right)\right]}\right\}^{-1}, 
\end{equation}
valid for $v<2\Delta_v$.
Using Eqs.\ \eqref{eq:S_approx} and \eqref{eq:rates}, we find for the
zero-frequency noise \cite{footnote:divergence}
\begin{align}
\label{eq:noise_telegraph}
S_\mathrm{m}(0)=&\,\frac{e^2\Gamma^2}{\omega_0}\gamma_{\mathrm{e}}^{-1}\frac{2\tilde T}{\Delta_v}
h^{-1}\left(\frac{v}{2\Delta_v}\right)
\nonumber\\
&\times\exp{\left(\frac{\Delta_v}{2\tilde T}\left[1-\frac{v}{\Delta_v}+\frac
12\left(\frac{v}{2\Delta_v}\right)^2\right]\right)}
\nonumber\\
&\times
\left\{1+2\exp{\left[\frac{\Delta_v}{4\tilde
T}\left(1-\frac{v}{\Delta_v}\right)\right]}\right\}^{-3}, 
\end{align}
where the function $h(z)$ is defined in Eq.\ \eqref{eq:h}.

At a bias voltage corresponding to the energy gap \eqref{eq:gap} ($v=\Delta_v$),
the results of Eqs.\ \eqref{eq:current_telegraph} and \eqref{eq:noise_telegraph}
simplify to yield the Fano factor
\begin{equation}
\label{eq:Fano_telegraph}
F_\mathrm{m}=\frac{16\Gamma}{3\omega_0}\gamma_\mathrm{e}^{-1}\frac{\tilde
T}{\Delta_v}\exp{\left(\frac{\Delta_v}{16\tilde T}\right)}. 
\end{equation}
Although this result is based on a simplified model and despite the fact that it
does not include the full nonequilibrium dynamics of the nanoresonator induced
by the charge fluctuations on the dot, it qualitatively captures
our main finding depicted in Fig.\ \ref{fig:main_result}. Indeed, as one approaches
the Euler instability from below, the gap $\Delta_v$ increases algebraically as
$\sim 1/|\delta|$ [cf.\ Eq.\ \eqref{eq:gap}], resulting in an exponential
increase of the Fano factor.

The results of Eqs.\ \eqref{eq:current_telegraph},
\eqref{eq:noise_telegraph} and \eqref{eq:Fano_telegraph} also apply for
compression forces far above the Euler instability
($\delta\gg\tilde\alpha^{1/3}$). \cite{footnote:v_eff} Since the energy gap
$\Delta_v$ is, in that case, half of the gap far below the instability
[$-\delta\gg\tilde\alpha^{1/3}$, cf.\ Eq.\ \eqref{eq:gap}], this explains the
asymmetry of the Fano factor for negative and positive $\delta$ in Fig.\
\ref{fig:main_result}. In the vicinity of the buckling instability
($|\delta|\ll\tilde\alpha^{1/3}$), the exponential dependence of the Fano factor
as a function of the gap \eqref{eq:gap} (which here scales with
$\tilde\alpha\ll1$ as $1/\tilde\alpha^{1/3}$) should stay qualitatively the same.
This results in a Fano factor which saturates at its maximal value at the Euler instability (cf.\
Fig.~\ref{fig:main_result}).

\begin{figure}[tb]
\includegraphics[width=\columnwidth]{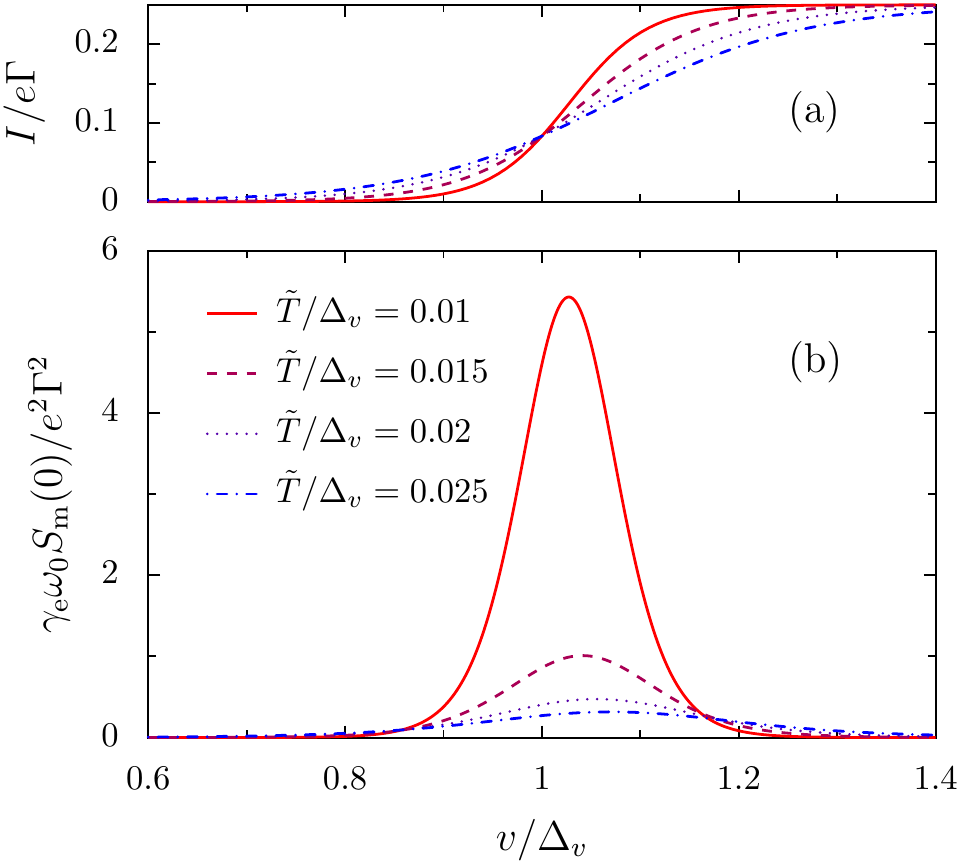}
\caption{\label{fig:telegraph}%
(Color online) 
(a) Current \eqref{eq:current_telegraph} and (b) zero-frequency 
noise \eqref{eq:noise_telegraph} as a function
of bias voltage for increasing
temperature as obtained from the telegraph-noise model.}
\end{figure}

The results of Eqs.\ \eqref{eq:current_telegraph} and \eqref{eq:noise_telegraph}
for the average current and the zero-frequency noise are shown in Figs.\
\ref{fig:telegraph}(a) and \ref{fig:telegraph}(b), respectively. As one can see from Fig.\ 
\ref{fig:telegraph}(b), our analytical results
capture the following trends for the mechanical noise: 
(i) it only depends on the
compression force $\delta$ through the ratio $\tilde T/\Delta_v$, 
(ii) the noise is maximal close to
$v=\Delta_v$ (a feature which has also been found in Ref.\
\onlinecite{pisto08_PRB}
in the case of fully coherent transport) and its maximum shifts towards higher bias voltages when one
increases the temperature, and (iii) the noise is inversely proportional to the extrinsic damping
constant (i.e., proportional to the quality factor).
In Sec.\ \ref{sec:numerics}, these features based on our simplified telegraph-noise model will be
confirmed and discussed further in the context of numerical calculations based on Eq.\
\eqref{eq:S_transformed}.

\subsection{Numerical results}
\label{sec:numerics}

\begin{figure}[b]
\includegraphics[width=\columnwidth]{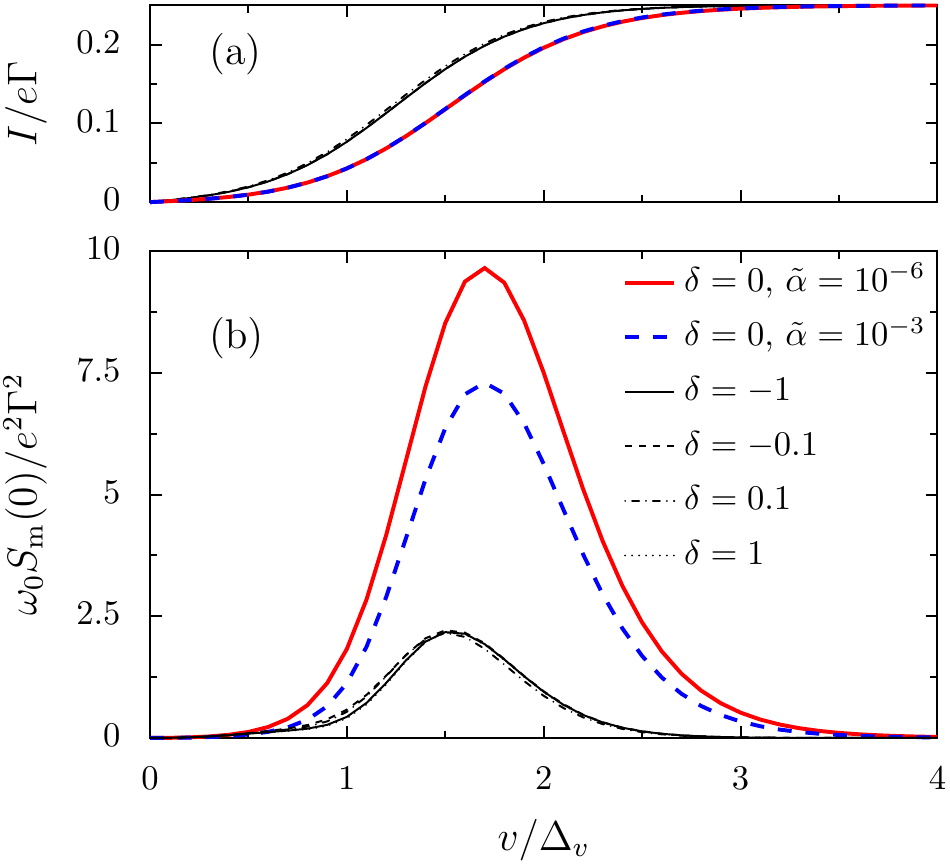}
\caption{\label{fig:scaling}%
(Color online) (a) Current and (b) zero-frequency noise
as a function of bias voltage 
for increasing values of the reduced compression force
$\delta$. In the figure, 
$v_\mathrm{g}=v_\mathrm{g}^\mathrm{min}$,
$\tilde T=\Delta_v/10$, $\gamma_\mathrm{e}=10^{-2}$,
$\omega_0/\Gamma=0$ (no current-induced fluctuations),
and $\tilde\alpha=10^{-6}$ for all lines except for the thick blue dashed one,
where $\tilde\alpha=10^{-3}$.}
\end{figure}

In Fig.\ \ref{fig:scaling}, we present numerical results for the average
current [Fig.\ \ref{fig:scaling}(a)] and the zero-frequency noise [Fig.\
\ref{fig:scaling}(b)] as a function of bias voltage, for a gate voltage
corresponding to the apex of the Coulomb diamond, see Eq.\ \eqref{eq:vg}.
As one can see from the figure, the noise has qualitatively the same behavior
far from the Euler instability [$|\delta|\gg\tilde\alpha^{1/3}$, see thin black
lines in Fig.\ \ref{fig:scaling}(b)] and at the instability [$\delta=0$, thick
lines in Fig.\ \ref{fig:scaling}(b)]. Remarkably, the noise [as well as the
current, see Fig.\ \ref{fig:scaling}(a) and Ref.\ \onlinecite{weick11_PRB}],
once plotted as a function of $v/\Delta_v$, and for the same value of $\tilde
T/\Delta_v$, is (almost) quantitatively the same for any $\delta$ far from the
instability [see thin black lines in Fig.\ \ref{fig:scaling}(b)]. 
The overall behavior of the current and noise in Fig.\ \ref{fig:scaling} when
the system is far from the Euler instability (see thin solid black lines in
Fig.\ \ref{fig:scaling}) can
be understood in terms of the telegraph-noise model detailed previously.
Indeed, Eq.\ \eqref{eq:noise_telegraph} shows that for
$\tilde T\ll\Delta_v$, the
noise is exponentially sensitive to the ratio $\Delta_v/\tilde T$ only,
thus explaining the scaling in Fig.\ \ref{fig:scaling}(b). 
In contrast, the above scaling does not apply at the mechanical instability
[see thick red and dashed blue lines in Fig.\ \ref{fig:scaling}(b)].
There, the noise is larger for smaller values of the anharmonicity parameter
$\tilde\alpha$, the latter determining the strength of the quartic correction in
the effective potential \eqref{eq:v_eff}.  

\begin{figure}[tb]
\includegraphics[width=\columnwidth]{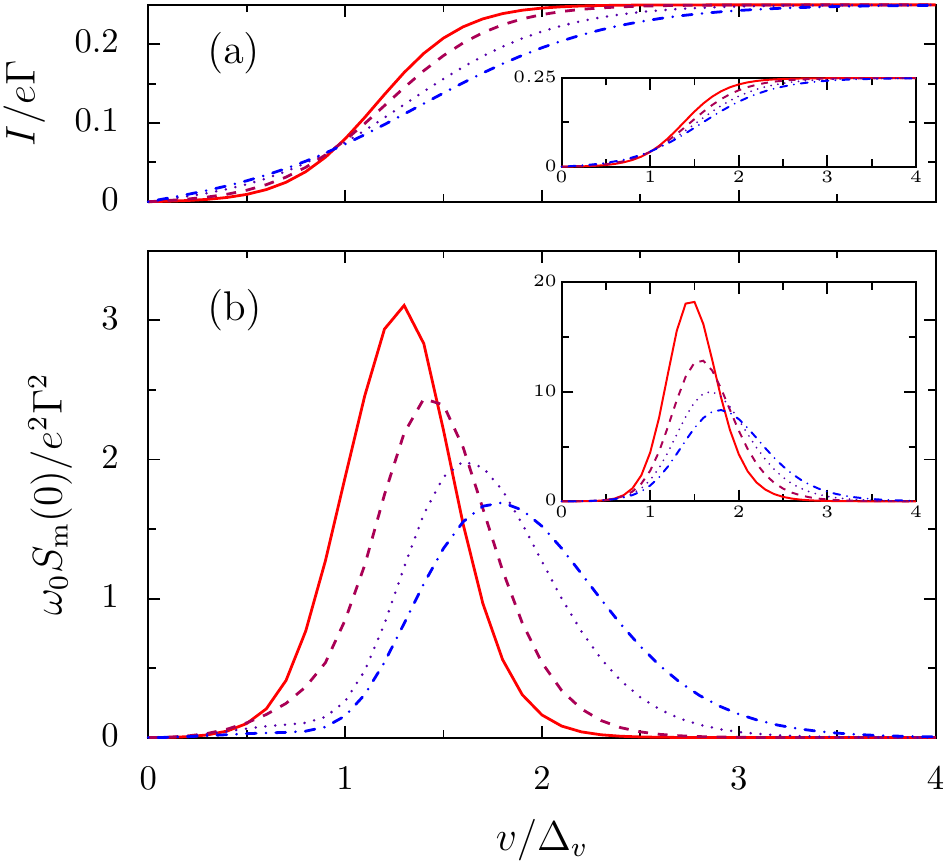}
\caption{\label{fig:temperature}%
(Color online) (a) Current and (b) zero-frequency noise
as a function of bias voltage
for increasing temperature $\tilde T=0.03$ (solid
line), $0.04$ (dashed line),
$0.06$ (dotted line), $0.08$ (dash-dotted line).
In the figure, $\delta=-1$, $v_\mathrm{g}=v_\mathrm{g}^\mathrm{min}$, $\gamma_\mathrm{e}=10^{-2}$,
$\omega_0/\Gamma=0$,
and $\tilde\alpha=10^{-6}$.
Insets: Same as the main figure for $\delta=0$.
In the insets, $\tilde T=2$ (solid line), $2.5$ (dashed line), $3$ (dotted line),
$3.5$ (dash-dotted line).}
\end{figure}

In Fig.\ \ref{fig:temperature}, we present the temperature dependence of both
the current and the zero-frequency noise far below the Euler instability [Figs.\
\ref{fig:temperature}(a) and \ref{fig:temperature}(b), respectively]  and at the
instability (see insets in Fig.\ \ref{fig:temperature}). As one can see from the
figure, the behavior of
these two quantities is qualitatively the same far from and at the buckling
instability. As temperature increases, the current blockade becomes less
pronounced [see Fig.\ \ref{fig:temperature}(a)] as
the system can explore more phase-space due to thermal fluctuations (for more
details, see Ref.\ \onlinecite{weick11_PRB}). Moreover, the maximum of the
zero-frequency noise decreases with temperature and gets shifted to larger
values of the bias voltage [see Fig.\ \ref{fig:temperature}(b)], a phenomenon which is captured by our analytical
estimate of the noise in Sec.\ \ref{sec:telegraph} (see Fig.\ \ref{fig:telegraph}). As temperature
increases, the probability to jump out of one minimum of the effective potential
(cf.\ Fig.\ \ref{fig:v_eff}) increases exponentially, such that the
associated telegraph noise decreases. 

\begin{figure}[tb]
\includegraphics[width=\columnwidth]{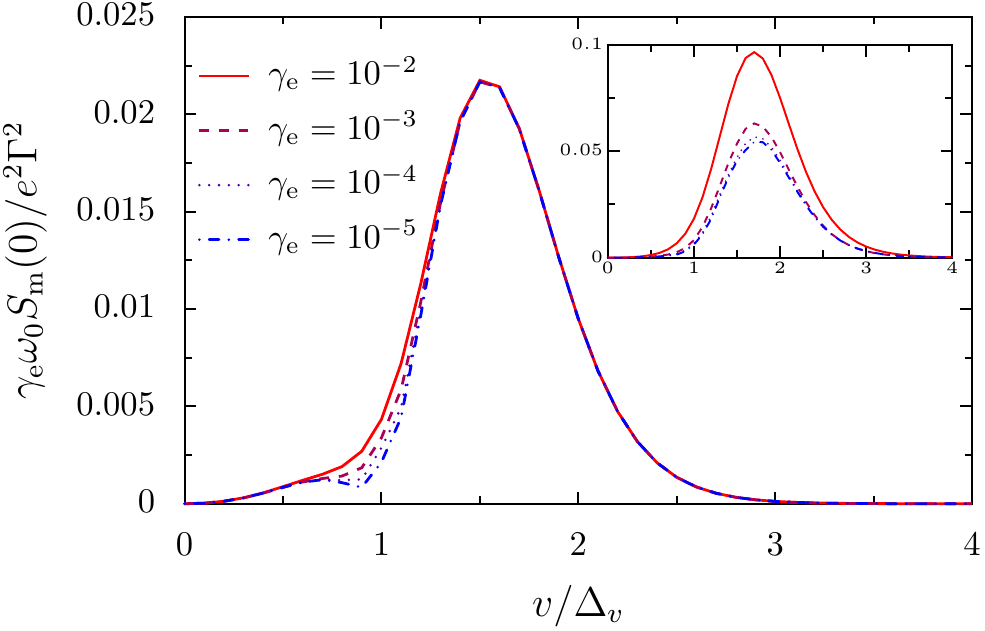}
\caption{\label{fig:gamma_e}%
(Color online) Zero-frequency noise
as a function of bias voltage 
for increasing values of the extrinsic damping constant
(or inverse quality factor) $\gamma_\mathrm{e}$.
In the figure, $\delta=-1$, $v_\mathrm{g}=v_\mathrm{g}^\mathrm{min}$, $\tilde T=\Delta_v/10$, 
$\omega_0/\Gamma=0$,
and $\tilde\alpha=10^{-6}$. Inset: Same as the main figure for $\delta=0$.}
\end{figure}

In Fig.\ \ref{fig:gamma_e}, we show numerical results for the zero-frequency
noise for various values of the extrinsic damping constant
$\gamma_\mathrm{e}=1/Q$
(or inverse quality factor) far below the instability (Fig.\ \ref{fig:gamma_e})
and at the Euler instability (inset in Fig.\ \ref{fig:gamma_e}). As one can see
from the main figure, the mechanically-induced noise scales almost perfectly as
$S_\mathrm{m}(0)\sim\gamma_\mathrm{e}^{-1}$ for compression forces far below the
instability. This is characteristic of telegraph
noise in the weak friction limit, \cite{krame40_Physica, hangg90_RMP} where the
system's energy varies only slowly
with time [cf.\ Appendix \ref{sec:rates} and Eq.\
\eqref{eq:noise_telegraph}]. 
Interestingly, in the case of molecular devices, the noise
is also much larger for unequilibrated (high-$Q$) vibrons. \cite{koch05_PRL, koch06_PRB}
At the instability (see the inset in Fig.\
\ref{fig:gamma_e}), the scaling with $\gamma_\mathrm{e}$ is only
approximate. \cite{footnote:gamma_e}

\begin{figure}[tb]
\includegraphics[width=\columnwidth]{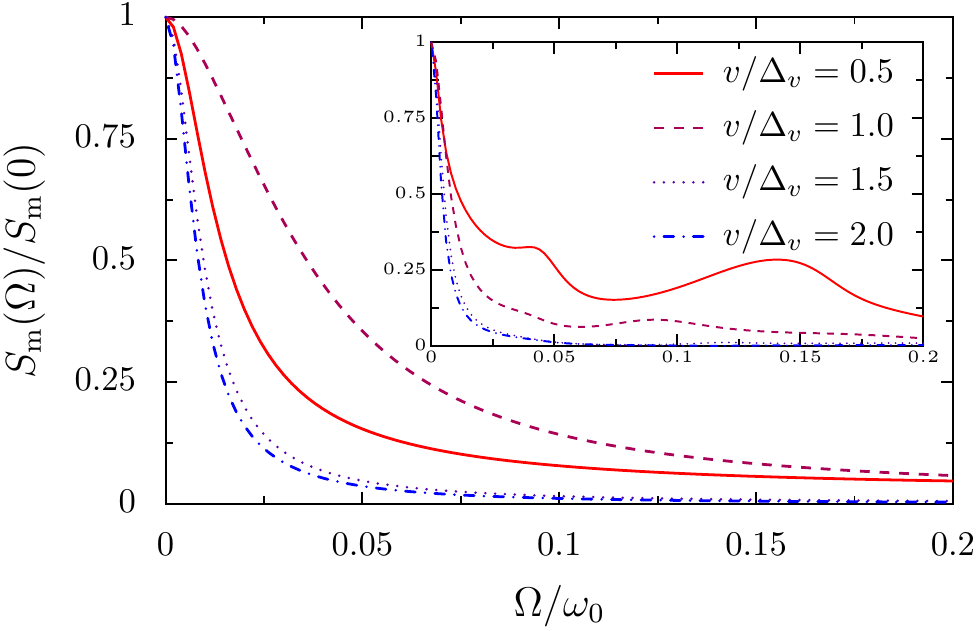}
\caption{\label{fig:frequency}%
(Color online) 
Frequency dependence of the mechanical noise for various values of the bias
voltage.  In the figure, $\delta=-1$, 
$v_\mathrm{g}=v_\mathrm{g}^\mathrm{min}$,
$\tilde
T=\Delta_v/10$, $\tilde\alpha=10^{-6}$, $\gamma_\mathrm{e}=10^{-2}$, and
$\omega_0/\Gamma=0$. (Inset) Same as the main
figure for $\delta=0$.}
\end{figure}

We conclude this section by computing the frequency dependence of the mechanical
noise \eqref{eq:S_transformed} far below (Fig.\ \ref{fig:frequency}) and at the
Euler instability (inset in Fig.\ \ref{fig:frequency}). Far from the
instability, the frequency dependence of the noise shows a $1/f^2$ dependence,
typical of telegraph noise [cf.\ Eq.\ \eqref{eq:S_approx}]. Notice that the
width of the Lorentzian shape of $S_\mathrm{m}(\Omega)$ depends on the bias
voltage through the transition rates for the system to enter or leave the
conducting minimum corresponding to an average occupation of the dot of $1/2$,
see Eqs.\ \eqref{eq:S_approx} and \eqref{eq:rates}.
At the Euler buckling instability, $S_\mathrm{m}(\Omega)$
also follows a $1/f^2$ behavior for $\Omega\ll\omega_0$, although additional
structures appear for larger frequencies and for certain bias voltages. 
\cite{footnote:frequency}

\section{Nonequilibrium fluctuations}
\label{sec:nonequilibrium}
We now investigate the mechanical noise in the presence of the current-induced 
fluctuations and dissipation, Eqs.\ \eqref{eq:d} and \eqref{eq:gamma}.

\begin{figure}[tb]
\includegraphics[width=\columnwidth]{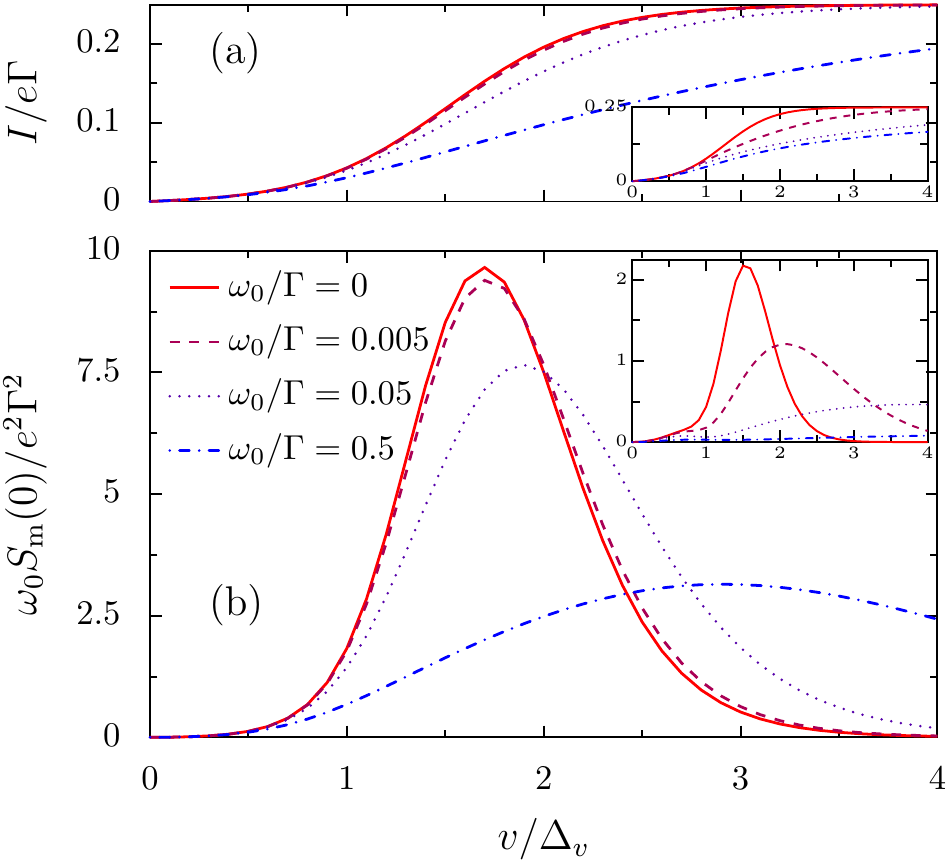}
\caption{\label{fig:nonequilibrium}%
(Color online) (a) Current and (b) zero-frequency noise
as a function of bias voltage 
for increasing values of the adiabaticity parameter
$\omega_0/\Gamma$. 
In the figure, $\delta=0$, $v_\mathrm{g}=v_\mathrm{g}^\mathrm{min}$, $\gamma_\mathrm{e}=10^{-2}$,
$\tilde T/\Delta_v=0.1$ and 
$\tilde\alpha=10^{-6}$. (Inset) Same as the main figure with
$\delta=-1$.}
\end{figure}

Our numerical results for the current and the zero-frequency noise are shown in
Figs.\ \ref{fig:nonequilibrium}(a) and \ref{fig:nonequilibrium}(b),
respectively, 
when the nanobeam is at the Euler instability (the insets in Fig.\
\ref{fig:nonequilibrium} consider the case $\delta=-1$).
As one can see from Fig.\ \ref{fig:nonequilibrium}, the effect of
an increasing adiabaticity parameter $\omega_0/\Gamma$, which controls the
strength of the current-induced fluctuations \eqref{eq:d} and dissipation
\eqref{eq:gamma}, is qualitatively similar to the effect of an increasing
temperature (cf.\ Fig.\ \ref{fig:temperature}). Indeed, the overall noise level
is reduced and the maximum of the noise is shifted towards larger bias voltages 
for increasing $\omega_0/\Gamma$ [see Fig.\ \ref{fig:nonequilibrium}(b)]. 
Moreover, the current blockade gets less pronounced for increasing
$\omega_0/\Gamma$ [see Fig.\ \ref{fig:nonequilibrium}(a)]. 

The results of Fig.\ \ref{fig:nonequilibrium} can be qualitatively understood by
defining an effective temperature \cite{weick11_PRB, weick11b_PRB}
\begin{equation}
\label{eq:T_eff}
\tilde T_\mathrm{eff}=\frac{\langle d\rangle/2+\gamma_\mathrm{e}\tilde
T}{\langle\gamma\rangle+\gamma_\mathrm{e}}
\end{equation}
in analogy with the fluctuation-dissipation theorem. \cite{zwanzig}
Here, $\langle d\rangle$ and $\langle\gamma\rangle$ are the averages over
phase-space of the current-induced fluctuations and dissipation, Eqs.\
\eqref{eq:d} and \eqref{eq:gamma}, respectively.
As shown in Ref.\ \onlinecite{weick11_PRB}, 
the effective temperature \eqref{eq:T_eff}
is approximately given by
$\tilde T_\mathrm{eff}\simeq\tilde
T+\frac{\omega_0/\Gamma}{4\gamma_\mathrm{e}}\Theta(v-\Delta_v)$, 
with $\Theta(z)$ the Heaviside step function.
This explains why both the current and the zero-frequency noise
are quite insensitive to the ratio $\omega_0/\Gamma$ for $v<\Delta_v$ and are
similar to the case $\omega_0/\Gamma=0$, i.e., the fully adiabatic limit.
On the contrary, for $v>\Delta_v$, 
the effective temperature increases with increasing $\omega_0/\Gamma$,
explaining the similarity of the behavior of the current and noise in Figs.\
\ref{fig:nonequilibrium} and \ref{fig:temperature}.

We conclude this section by noticing that we have numerically checked that the frequency dependence of the noise
power spectrum also follows a $1/f^2$ behavior when one takes into account
current-induced fluctuations, similar to Fig.\ \ref{fig:frequency}.
This confirms that the noise is dominated by a telegraph noise at
low-enough temperatures even in presence of nonequilibrium fluctuations.

\section{Conclusion}
\label{sec:ccl}
We have investigated the current noise in nanoelectromechanical
systems close to a continuous mechanical instability. Specifically, we have
considered the paradigmatic Euler buckling instability in suspended
single-electron transistors which are capacitively coupled to a gate electrode. 
We have predicted a drastic enhancement of the current noise when
the nanobeam supporting the quantum dot is brought to the Euler instability,
resulting in very large Fano factors that are well above the Poisson limit. This
exponential enhancement at the buckling instability is directly related to the
(algebraic) enhancement of the current blockade predicted in Ref.\
\onlinecite{weick11_PRB}. We developed a rather detailed picture of the
underlying physics in terms of a telegraph-noise model.
While such large Fano factors may make
observation of the low-bias current blockade more challenging, the large noise
levels predicted in this work
would serve also as a clear experimental signature  of the
interplay between electronic and mechanical degrees of freedom in NEMS (e.g.,
carbon nanotubes \cite{onac06_PRL}) close to 
continuous mechanical instabilities.

\begin{acknowledgments}
We thank Niels Bode for useful discussions. JB and FvO acknowledge 
the Deutsche Forschungsgemeinschaft through Sonderforschungsbereich 658 for
financial support.
FP acknowledges support from the French ANR through contract QNM No.\ 0404 01.
\end{acknowledgments}

\appendix
\section{Shot noise}
\label{sec:shot}
In this appendix, we briefly comment on how the shot noise properties of the
setup considered in this paper (see Fig.~\ref{fig:setup}) are influenced by the
coupling \eqref{eq:H_c} between the charge on the quantum dot and the mechanical degree of freedom. 

Within the adiabatic approach presented in Sec.\
\ref{sec:model}, the shot noise reads
\begin{equation}
\label{eq:S_sh}
S_\mathrm{sh}(\Omega)=\int \mathrm{d}x\mathrm{d}p\;\mathcal{P}_\mathrm{st}(x,
p)\mathcal{S}_\mathrm{sh}(\Omega, x), 
\end{equation}
where $\mathcal{S}_\mathrm{sh}(\Omega, x)$ is the quasistationary shot noise
for fixed position $x$. Its zero-frequency limit ($\Omega\ll\Gamma$) 
\cite{blant00_PhysRep} reads
in the sequential-tunneling regime ($\hbar\Gamma \ll k_\mathrm{B}T$)
\begin{equation}
\label{eq:S_sh(0,x)}
\mathcal{S}_\mathrm{sh}(0, x)=
\frac{e^2\Gamma}{4}
\left[
f_\mathrm{L}(x)+f_\mathrm{R}(x)
\right]
\left[2-
f_\mathrm{L}(x)-f_\mathrm{R}(x)
\right]
\end{equation}
where $f_\mathrm{L/R}(x)$ are the Fermi factors defined in
Eq.~\eqref{eq:Fermi}.

\begin{figure}[tb]
\includegraphics[width=\columnwidth]{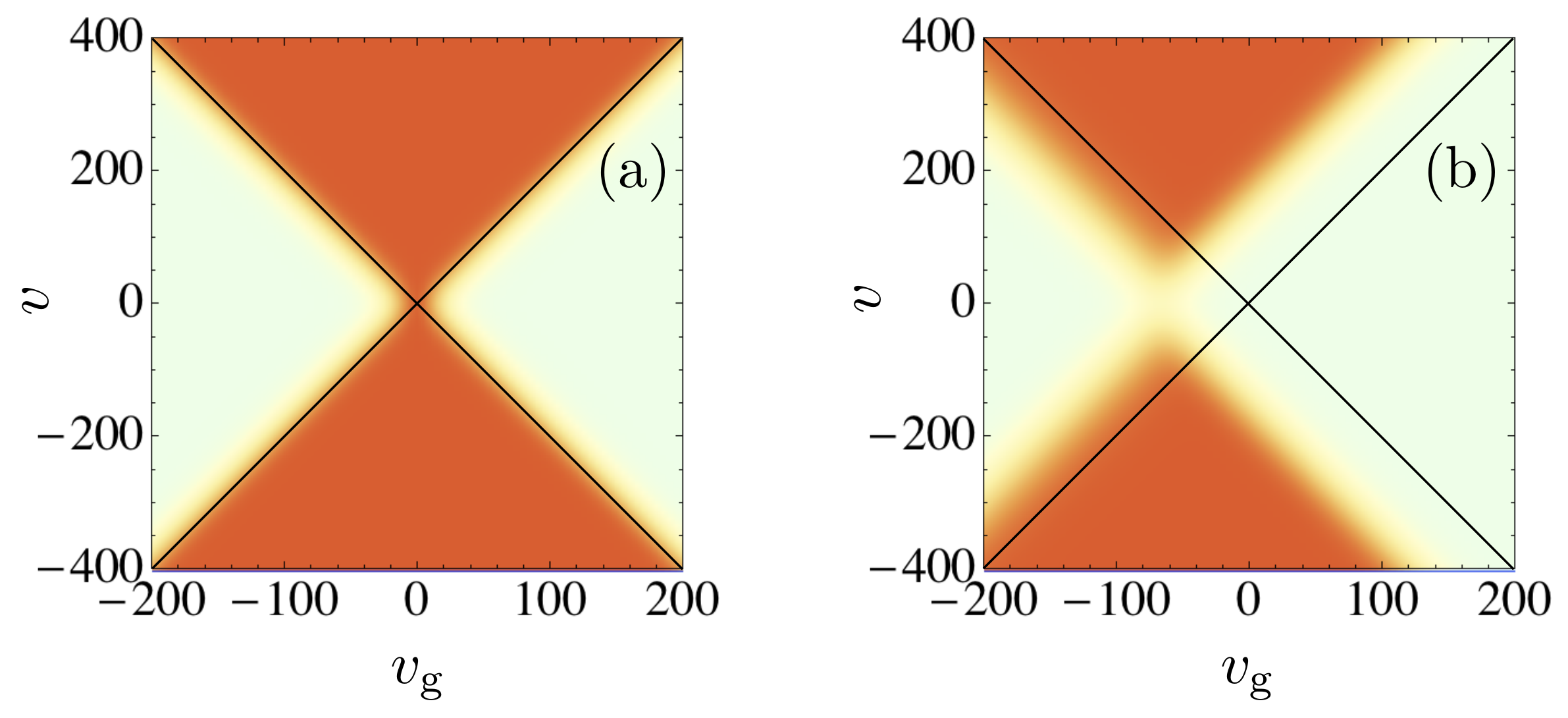}
\caption{\label{fig:shot_noise}%
(Color online) Zero-frequency shot noise as a function of 
bias $v$ and gate voltage
$v_\mathrm{g}$
(a) for vanishing compression force $F=0$ and (b) at the Euler
buckling instability where $F=F_\mathrm{c}$. 
The parameters used in the figure are
$\tilde T=10$, 
$\tilde\alpha=10^{-6}$,
$\omega_0/\Gamma=10^{-2}$, 
and $\gamma_\mathrm{e}=1/Q=10^{-2}$. 
Color scale: 
white and red (light gray) regions correspond to $S_\mathrm{sh}(0)=0$
and $e^2\Gamma/4$, respectively.}
\end{figure}

In the strictly adiabatic limit $\omega_0/\Gamma\rightarrow 0$, the
current-induced fluctuation and damping coefficients are both vanishing
[see Eqs.~\eqref{eq:d} and \eqref{eq:gamma}], 
such that the stationary
probability distribution function corresponding to the Fokker-Planck equation
\eqref{eq:FP_scaled} is a Boltzmann distribution at temperature $\tilde T$.
Assuming zero temperature, 
the (zero-frequency) shot noise \eqref{eq:S_sh}
reduces to $S_\mathrm{sh}(0)=\mathcal{S}_\mathrm{sh}(0, x_\mathrm{m})$, 
with $x_\mathrm{m}$ the global minimum
of the effective potential \eqref{eq:v_eff}.
\cite{weick11_PRB}
Thus, the
shot noise observes the same behavior as the mean-field current discussed in
Ref.~\onlinecite{weick11_PRB}: In the $v$-$v_\mathrm{g}$ plane, 
the shot noise has the
same structure as the Coulomb diamonds delimited by slopes $v\sim \pm 2v_\mathrm{g}$,
the apex of these diamonds being given by Eqs.~\eqref{eq:gap} and \eqref{eq:vg}.
The zero-frequency shot noise thus takes the value
$S_\mathrm{sh}(0)=e^2\Gamma/4=e|I|$ in the conducting regions of the
$v$-$v_\mathrm{g}$ plane, with a corresponding Fano factor
$F_\mathrm{sh}=S_\mathrm{sh}(0)/2e|I|$ of $1/2$, typical for a
single-level quantum dot symmetrically coupled to source and drain leads in the
sequential-tunneling regime. \cite{blant00_PhysRep}
The behavior of the mean-field zero-frequency shot noise as a function of the reduced
compression force $\delta$ can thus readily be deduced from the behavior of the
mean-field current shown in Fig.~4 of Ref.~\onlinecite{weick11_PRB} (see also
Fig.\ 2 in Ref.~\onlinecite{weick11b_PRB}).

Including thermal as well as current-induced fluctuations by solving for the
stationary solution of
the Fokker-Planck equation \eqref{eq:FP_scaled} numerically and computing the
shot noise with the help of Eq.~\eqref{eq:S_sh} leads to
qualitatively the same effects as for the average current. This is exemplified in
Fig.~\ref{fig:shot_noise}, where fluctuations lead to a softening of the borders
in the $v$-$v_\mathrm{g}$ plane 
delimiting the regions with a finite shot noise $S_\mathrm{sh}\simeq
e^2\Gamma/4$.

\section{Transition rates}
\label{sec:rates}
The transition rates $\Gamma_\mathrm{in}$ and $\Gamma_\mathrm{out}$ entering 
our approximate expressions for the average current \eqref{eq:I_approx} and 
the noise power spectrum \eqref{eq:S_approx} can be
calculated using Kramers reaction rate theory. \cite{krame40_Physica, 
hangg90_RMP, zwanzig}
In the weak damping regime, which is the experimentally relevant one for carbon
nanotube-based resonators that can present very high quality factors,
\cite{steel09_Science, lassa09_Science} the
escape rate from the minimum located at $x_i$ is given by
\begin{equation}
\label{eq:escape_rate}
k_{x_i}=\gamma_\mathrm{e}\frac{\omega_0\sqrt{v_\mathrm{eff}''(x_i)}}{2\pi\tilde T}
S_0(v_{\mathrm{b},i})\exp{\left(-\frac{v_{\mathrm{b},i}}{\tilde T}\right)}, 
\end{equation}
where 
$S_0(v_{\mathrm{b},i})$ 
is the abbreviated action 
at the barrier
top, whose energy, seen from $x_i$, is denoted by $v_{\mathrm{b},i}$. 
Notice that Eq.\ \eqref{eq:escape_rate} is valid as long as $\tilde T\ll v_{\mathrm{b},i}$ and in the weak damping regime, i.e.,
$\gamma_\mathrm{e}S_0(v_{\mathrm{b},i})\ll \tilde T$. \cite{hangg90_RMP}
With Eq.\ \eqref{eq:escape_rate}, and taking into account the probability that
the system thermalizes in the well it jumps to, \cite{hangg90_RMP} we find for the transition rates 
\begin{subequations}
\label{eq:rates}
\begin{align}
\Gamma_\mathrm{in}&=\gamma_\mathrm{e}\omega_0\frac{\Delta_v}{4\tilde T}
h\left(\frac{v}{2\Delta_v}\right)
\exp{\left(-\frac{\Delta_v}{4\tilde
T}\left[1-\frac{v}{2\Delta_v}\right]^2\right)}, 
\\
\Gamma_\mathrm{out}&=\gamma_\mathrm{e}\omega_0\frac{\Delta_v}{2\tilde T}
h\left(\frac{v}{2\Delta_v}\right)
\exp{\left(-\frac{\Delta_v}{4\tilde
T}\left[\frac{v}{2\Delta_v}\right]^2\right)},
\end{align}
\end{subequations}
with
\begin{equation}
\label{eq:h}
h(z)=\frac{z^2(1-z)^2}{z^2+2(1-z)^2}.
\end{equation}



\begin{thebibliography}{}
\bibitem{footnote:v_eff}
Far above the Euler instability ($\delta\gg\tilde\alpha^{1/3}$), the effective
potential has, close to $x=-\sqrt{\delta/\tilde\alpha}$, approximately the same form as for
$-\delta\gg\tilde\alpha^{1/3}$, cf.\ Fig.\ \ref{fig:v_eff}(c).
\bibitem{footnote:numerics}
The stationary solution of the Fokker-Planck equation \eqref{eq:FP_scaled} is
obtained by discretizing phase-space over a finite grid using vanishing boundary
conditions and taking advantage of the normalization condition 
$\int\mathrm{d}x\mathrm{d}p\,\mathcal{P}_\mathrm{st}(x, p)=1$.
\bibitem{footnote:shuttle}
Note that, in the somewhat different context of the quantum shuttle, large Fano
factors have been predicted in Refs.\ \onlinecite{flind05_EPL} and
\onlinecite{donar05_NJP} in the so-called ``coexistence regime", which were
interpreted in terms of telegraph noise.
\bibitem{footnote:divergence}
The apparent divergence of this result for $v=0$ and $v=2\Delta_v$
is unphysical as these bias voltages are out of the range of validity 
of the transition rates \eqref{eq:rates} entering Eq.\
\eqref{eq:S_approx} (cf.\ the weak friction limit).
\bibitem{footnote:gamma_e}
As we have checked, for the largest values of $\gamma_\mathrm{e}$
used in the inset in Fig.\ \ref{fig:gamma_e}, 
the weak-friction limit used in  Appendix \ref{sec:rates} of
Kramers reaction rate theory is no longer strictly valid, such that the scaling
is only approximate.
\bibitem{footnote:frequency}
These features
cannot be explained by our telegraph-noise model presented above.
This might be due, e.g., to the asymmetry of the effective
potential at $\delta=0$ [see Fig.\ \ref{fig:v_eff}(f)] which is not incorporated
in our analytical model, the latter being in principle only valid far below the
instability.
\end{thebibliography}
\end{document}